# The recurrent nova T CrB had prior eruptions observed near December 1787 and October 1217 AD

**Bradley E. Schaefer**
Louisiana State University, USA

## Abstract

The famous recurrent nova (RN) T Coronae Borealis (T CrB) has had observed eruptions peaking at a visual magnitude of 2.0 in the years 1866 and 1946, while a third eruption is now expected for the year $2024.4 \pm 0.3$. Each RN has very similar light curves of eruptions that come with a fairly even-spacing in time, for which T CrB has a recurrence timescale near 80 years. So it is reasonable to look backwards in time for prior eruptions, around 1786, and so on back. I have investigated two long-lost suggestions that T CrB was seen in eruption in the years 1217 and 1787. (1) In a catalog published in 1789, the Reverend Francis Wollaston reports an astrometric position for a star that is exactly on top of T CrB. From his letters, these observations were made on at least four occasions with both a large and small telescope, within a few days before 1787 December 28. Wollaston's limiting magnitude for his astrometry is near 7.8 mag, so T CrB would have to have been in eruption. Wollaston's quoted coordinates are not from the nearest star above his limit, HD 143707, because any mistake that places the error box on top of T CrB has much too small a probability to be acceptable. With other transients strongly rejected, the only way that Wollaston could get the coordinates was to have measured the coordinates of T CrB itself during an eruption. (2) The 1217 event has an eyewitness report written by Abbott Burchard of Upsberg as a fast-rising stellar point-source ("stella") in Corona Borealis that "shone with great light," lasted for "many days," and was ascribed as being a "wonderful sign." This event cannot be a report of a comet, because Burchard used the term for a star ("stella") and not for a comet, and because Burchard had the omen being very positive, with such being impossible for comets that are universally the worst of omens. The reported event is just as expected for a prior eruption of T CrB, and all other possibilities are strongly rejected, so the case for the 1217 eruption of T CrB is strong.

**Corresponding author:**
Bradley E. Schaefer, Department of Physics and Astronomy, Louisiana State University, Baton Rouge, LA 70803-0001, USA.
Email: schaefer@lsu.edu

## Introduction

T Coronae Borealis[1] (T CrB) is a recurrent nova[2] (RN) with observed eruptions in the years 1866 and 1946. In 1866, T CrB was the first well-observed and widely-observed nova event, and was the first nova observed spectroscopically.[3] The nova peaked at visual magnitude 2.0, and has been the brightest nova in Earth's skies since 1942 for CP Pup.[4] In quiescence, T CrB is by far the brightest of all nova systems, at a visual magnitude of 9.8. In 2015, with the onset of its unique and characteristic high-state, many workers have realized that T CrB will soon undergo another nova event, estimated for the year $2025.5 \pm 1.3$. In June 2023, I announced the start of the long-anticipated pre-eruption dip starting in late March,[5] for a possibly-improved eruption date expected for $2024.4 \pm 0.3$. This bright nova has attracted a variety of evocative storylines.[6] With the upcoming eruption, T CrB will again come to prominence for the public, journalists, and astronomers.

Successive eruptions from each individual RN system (including T CrB) are identical (to within small measurement errors) to each other both spectroscopically and photometrically.[7] Further, their relative timings are modestly regular, with variations of 4%–40% in the inter-eruption intervals.[8] With eruptions in 1946 and 1866, T CrB has a recurrence time of approximately 80 years. Nothing concerning T CrB eruptions is changing substantially on any timescale under a million years, so there is every expectation that the nova has been erupting to 2.0 mag at ~80 years intervals going far into the past. With T CrB as one of the all-time brightest novae, shining brightly every eighty or so years, we would expect that prior eruptions would have been discovered by many observers in many centuries throughout the northern hemisphere over the last two millenia. The task is to find whether any of these inevitable discoveries resulted in a report surviving into modern times, and to test whether any such reports are convincing.

Immediately upon the discovery of the 1866 eruption. Sir John F. W. Herschel[9] claimed to have seen T CrB in outburst on 1842 June 9. Herschel's evidence consisted only of a drawing he made where he was charting all stars visible in the entire sky.[10] Herschel's 1842 chart[11] is now only known from the accurate "fair copy" he made in 1866,[12] and this shows that the star of interest is actually about 1° from the position of T CrB, and is closely matched with the position of the ordinary G8 star HD 144287 (BD + 25°3020). With this, I have proven that in the year 1842 John Herschel did *not* see T CrB.[13]

The secondary literature contains *two* further claims for observations of pre-1866 T CrB eruptions[14] in the years 1217 and near 1788 Both of these claims are only of a few sentences, presented only as weak suggestions, and both have been entirely ignored in all later discussions and publications. Both suggestions have been entirely forgotten, except by R. Webbink, where these citations have been included in his exhaustive bibliography of pre-1977 papers and manuscripts touching on T CrB.[15] With ordinary curiosity, I became interested in Webbink's two citations that claim old eruptions. On checking with the primary sources, I found that both claims are reasonable and deserving of full investigation. This paper is my report on both claims.

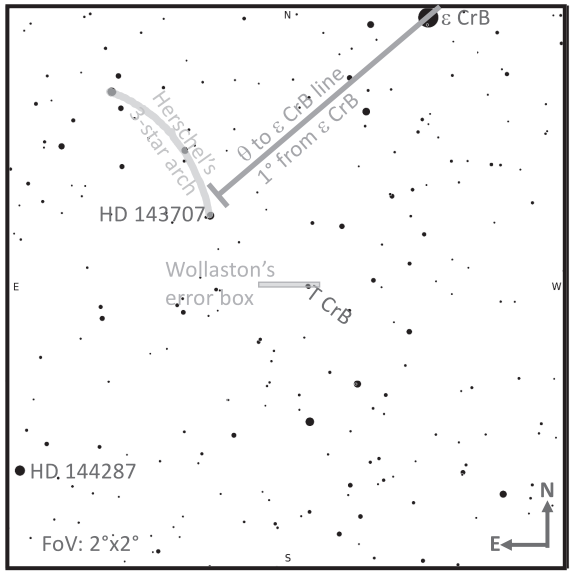

**Figure 1.** Star chart for the area around T CrB. T CrB itself is the small dot in the center. Wollaston's error box from the entry in his 1789 catalog is shown as a long thin green rectangle in the center. The height and width of the error box is defined by the precision for Wollaston's quoted coordinates, $\pm 0.11°$ in RA and $\pm 0.008°$ in declination. Importantly, the Wollaston error box covers the T CrB position. Herschel's double star V 75 was described by Herschel (in his second double star catalog) to be 1° from ε CrB along a extension of the oblique line (orange line in the upper middle) from θ CrB (far outside the figure to the upper-right) to ε CrB, with the position 1° from ε CrB along this extension represented by the bar at the end of the orange line. Herschel's description further states that V 75 is the westernmost star in an arch of three stars, with this identified by the dark-yellow arc in the upper left. Both of these description point to Herschel's double star V 75 being the star with the modern name of HD 143707. The chart shows the faint member of Herschel's double as a small dot on the edge of the larger dot for the primary, off at a position angle near 106° east of north. So there is no doubt about the position of Herschel's double star V 75. Importantly, this position is 0.39° from the Wollaston's error box from his 1789 catalog. Given either Herschel's positional uncertainties or Wollaston's positional uncertainties, the probability is unacceptably low that any ordinary error for the position of HD 143707 would result in such a large deviation that happens to land on top of T CrB. This is the strong argument that Wollaston's 1789 catalog coordinates are not from V 75. The field of view is square with sides $2° \times 2°$, north is up and east is to the left, and the chart shows stars down to visual magnitude of 13.0. We see that there are no stars near T CrB that were bright enough for Wollaston to measure, until we get to HD 143707. So if Wollaston were seeking to measure V 75 at a time when T CrB is in eruption, then it would be easy for him to mistakenly measure the coordinates of T CrB instead. Indeed, Wollaston's coordinates match those of T CrB to good precision, so this is a very strong case that Wollaston measured T CrB at a time when it was brighter than his effective limiting magnitude of 7.8 mag. That is, Wollaston's coordinates are an eye-witness report of T CrB in eruption in the year 1788 or soon before.

**Table 1.** Star positions and their agreement with the star in Wollaston's catalog.

| Position | RA | Declination | Comparison |
|---|---|---|---|
| | J2000 | J2000 | |
| Wollaston's 1789 catalog | 239.95° ± 0.11° | 25.912° ± 0.008° | What is source of these coordinates? |
| T CrB | 239.876° | 25.920° | Good agreement with Wollaston's catalog entry |
| V 75: Herschel's descriptive | 240.23° | 26.24° | OK position for V 75 |
| V 75: HD 143707 | 240.266° | 26.172° | Far from coordinates of Wollaston's catalog entry |
| V 75: Wollaston 1784 chart | 240.27° | 26.17° | Good position for V 75 |
| HD 144287 | 241.015° | 25.255° | J. Herschel's 1842 star |

## The nova of December 1787 AD

In 1789, English astronomer Francis Wollaston (1731–1815) published a comprehensive catalog of all bright and interesting stars.[16] This was done by compiling many extant catalogs, including Flamstead's *British Catalogue* of stars, Messier's catalog of nebulae and clusters, William Herschel's catalog of double stars, and a dozen more catalogs. Wollaston's goal was to assemble the many disparate and conflicting catalogs, and to process all the coordinates to a single common epoch of 1790.0.

Wollaston's catalog reports a star at the position of T CrB,[17] and there are no normal stars brighter than 13 mag anywhere near to this position. Figure 1 provides a star chart for showing the relationships between the position for this star, T CrB, and other stars of relevance, while Table 1 collects together the J2000 coordinates of all the relevant star positions. The positional coincidence to T CrB is the same as the measurement precision, so the lack of any nearby stars makes for a low-probability of the star being a chance error of some type, hence making a good case that there is a causal connection from Wollaston's star to T CrB. The star with Wollaston's coordinates must have been viewed by a highly experienced observer with a telescope for some substantial length of time, so it is not a comet or asteroid, while other classes of variable stars are strongly rejected by their absence in modern deep surveys, so the only possible source for a transient stellar source at that location is T CrB itself. No magnitude is quoted by Wollaston, but the catalog limit is roughly a visual magnitude of 7.8, so the report must have been made at a time when T CrB was in eruption. Another connection to T CrB is that the recurrent nova had an eruption close to the year 1786 (=1866−80), so a T CrB eruption at this time is strongly expected. The T CrB hypothesis for Wollaston's star is very strong.

The entry for the star at T CrB's position appears in the catalog, ordered by right ascension (RA), for the zone with 63° north polar distance. The entry consists of three measures for the positional coordinates plus one relatively long descriptive text. The first measure in the entry is the RA listed as "±237.45," which is 237°45′, or 237.75°.

The ± symbol was used by Wollaston for all the stars with the information taken only from Herschel's double star catalog.[18] The second entry for the star's RA is "±51," with this being the value in time units, where the hour is implied to be 15 as given by the entries higher in the table. The second RA is thus 15 hours 51 minutes, which equals 237.75°, so the two quoted RA values are exactly equal. The RA is only reported to the nearest minute of time, so the precision of the measure is ±0.5 minute of time in RA, which equals ±6.7 arc-minutes or ±0.11°. The third number in the star's entry is "±29," which is the north polar distance of the star, with the degree component being 63°, so the north polar distance is 63°29′, or a declination of +26.516°. The precision of the declination is ±0.5 arc-minute, which equals ±0.008°. So the first three numbers in the entry give the coordinates of the star in 1790 coordinates as an RA of 237.75° ± 0.11° and a declination of +26.516° ± 0.008°. This reported precision gives an error box that is long and thin with dimensions 0.22° × 0.016°. I have processed these coordinates to the now-standard epoch of J2000 (see Table 1). This is to be compared to the modern J2000 position of T CrB itself (see Table 1). The error box is shown on the sky chart in Figure 1.

The last part of the entry for the star at the T CrB position is a shorthand description in the last column. The descriptive entry with the original typeface is:

"double (Herf. V. 75) v. v. uneq. . . dift.41".12'". . . . pof. 16° S. f. It is really quadruple, for the fmall Star is double, and there is a ftill fmaller at about 40°. S. p. the fmall ones."

Here is a "translation," with the jargon and shorthand expanded to modern English text:

"This entry is for the double star that appears in W. Herschel's [second] double star catalog, for star numbered 75 in Herschel's list for class V double stars (those with 30-60 arc-second separations). The two stars are very-very-unequal in brightness, with this corresponding to perhaps 2-5 mags. The fainter star of the double appears 41.2 arc-seconds from the brighter, with a position angle of 16° south of east, or 106° from north towards the east. [The preceding description is Wollaston's usual direct copying of the information from Herschel's double star catalog.] This star is really a quadruple star, i.e., four stars all close together, with the fainter star being itself double, plus there is a still fainter star towards the southwest of the fainter double, with a position angle of 230° from north through the east."

The first part of this description is a faithful and stylized reproduction of information from Herschel's double star catalog.[19] This star has a confident modern indentification as HD 143707. The coordinates are in Table 1 and Herschel's double star is indentified in Figure 1. The critical point from this is that HD 143707 is far from the coordinates quoted in Wollaston's catalog. To be specific, HD 143707 is 0.39° to the northeast of the coordinates quoted by Wollaston's catalog. This disagreement makes a good case that Wollaston's coordinates do not refer to HD 143707 or V 75.

The second part of the catalog's entry is to give a further description of the star. This added information is the only time in the catalog where Wollaston added any extra information for any star only cataloged by Herschel. So already we know that there is something special about this entry. This added information is not from anywhere in Herschel's catalog or in his logbooks.[20] Nor is it a mistaken duplication from some other Herschel double star, nor does the quadruple star description match any other stars in the T CrB area of the sky. I have not found any source for this extra added description. An important point is that the star described in the added notes is not HD 143707 or T CrB. This one entry is unique in Wollaston's catalog as being a mix-up that reports on multiple stars.

Wollaston's catalog entry is somehow combining observations of three separate stars[21]; a star at the coordinates of T CrB, Herschel's double star HD 143707, and a third unknown quadruple star. The coordinates came from some star up in the sky that was then brighter than 7.8 mag or so.

The T CrB hypothesis is that someone, likely Wollaston himself,[22] observed T CrB in the tail of an eruption in the year 1788 or soon before, measured the coordinates, and included these in the catalog as mistaken for Herschel's double star. In such a case, T CrB would appear brighter than HD 143707 (because HD 143707 is at the limiting magnitude for Wollaston's astrometry) so Wollaston would have measured the coordinates of the nova and used these coordinates in his catalog, in association with a shorthand version of Herschel's description of HD 143707. Wollaston added on the further mix-up by including a description of an unknown quadruple star for the last part of the catalog entry. With three stars going into this one entry, there must have been some special circumstances, for which the obvious possibility is confusion caused by the nova itself.

Within the T CrB hypothesis, I can likely pinpoint the observer and the date of the observations. Herschel first recorded HD 143707 as a double star (later cataloged as V 75), on 1782 July 18, and his observing logbooks show that he only observed this on three nights.[23] Herschel recorded the position of V 75 only as a verbal description star-hopping along an oblique line from ε CrB,[24] with this giving a position close to HD 143707 (see Figure 1 and Table 1). Herschel's catalog was presented to the Royal Society[25] on 1784 December 9. A thorough search of Herschel's logbooks shows that he never revisited this double.[26] Starting in 1783, Wollaston worked out improvements in the astrometric measurements of stars, and counseled that astronomers should undertake their own detailed measurements of all the stars in some constellation, as a check on earlier catalogs. As an example of his method, Wollaston chose Corona Borealis, where he measured the positions of 126 stars, visiting each four or more times, with both a large and a small telescope. Wollaston's astrometric catalog of CrB[27] was presented to the Royal Society on 1784 February 5. Wollaston's paper only displays his astrometry as a chart of the CrB area, recording stars with an average limiting magnitude of 7.8 mag. The star HD 143707 is accurately plotted (see Table 1 for the mapped coordinates), and there is no star anywhere near the T CrB position. So the coordinates that appeared in Wollaston's 1789 catalog can only have been measured after 1784 February. In May of 1785, Wollaston sent two letters to Herschel talking about his measures of double stars η CrB and δ CrB.[28] In a letter to Herschel dated 1787 Dec 28, Wollaston complains "I find a difficulty in ascertaining the places of some of your Double stars; particularly several of those in your Second Catalog, whose positions are noted down in an oblique line from some known star."[29] There is only one such star (in Herschel's second catalog, in CrB, placed with an oblique line from some known star), so Wollaston must be referring to the star he thought to be V 75. This letter shows that Wollaston was getting confused because his coordinates for V 75 did not agree with Herschel's descriptive position. Such confusion would naturally arise if Wollaston was actually measuring the coordinates of T CrB in outburst, and not the coordinates of V 75 as he supposed. To be above Wollaston's limiting magnitude, the star at T CrB's position would have to be brighter than 7.8 mag, and on at least four occasions. This means that T CrB must have been in eruption, and this would put the peak of the eruption sometime around 1787 December 20. Wollaston sent his catalog to the printers[30] in September or October of 1788. So Wollaston was an eye-witness of the T CrB eruption, and he measured the coordinates of T CrB in outburst around Christmastime in 1787.

We must examine all possible alternatives for the source of the coordinates in question:

**(A)** The T CrB hypothesis is that Wollaston observed the coordinates of T CrB in the tail of an eruption, and used these coordinates in his catalog for the star he thought was V 75.

**(B)** The HD hypothesis is that the coordinates are those of HD 143707 (V 75), just as Wollaston claims, and that a very-low-probability mistake made those coordinates match the position of T CrB.

**(C)** A faint transient could be a passing comet. This possibility has no real chance, partly because Wollaston was a highly experienced observer who had already published his own comet observations,[31] and he was using both large and small telescopes in a very dark site, so that the inevitable coma would be easily apparent. Further, he measured its astrometric position on at least four occasions, so the comet's motion would have been obvious.

**(D)** A faint transient could be a passing asteroid, and this gets around the comet-problem of the lack of any coma. But we still have Wollaston making astrometric measures on at least four occasions, so any asteroid motion would be obvious. Further, with an ecliptic latitude of $+45°$, no asteroid can ever get brighter than Wollaston's limit of 7.8 mag.

**(E)** Could the stellar transient seen by Wollaston be a supernova? No, because any supernova that gets somewhat brighter than 7.8 mag would have to be at 2.7 megaparsecs or somewhat closer, and would be in intergalactic space, more than 2.7 megaparsecs or so from any birth galaxy.[32] Further, any such supernova would have a young remnant that would be very bright in X-ray and radio, being discovered long ago along the line of sight to T CrB.

**(F)** Could Wollaston have sighted some eruptive variable star? Superflare stars and flare stars have too low an amplitude and too short a duration. Any dwarf nova that could reach 7.8 mag or brighter would have a relatively very-bright counterpart discovered long ago. Shell stars have events with durations of many years and only a modest amplitude. Gamma-ray bursts in the optical are always much too short in duration to allow Wollaston to make his required four observations, and there are no host galaxies near enough in the area. Wollaston could not have measured his coordinates for some other non-TCrB nova, because any quiescent counterpart would have already been discovered and because the probability of there being a second galactic nova along the T CrB line of sight is too small to accept.[33]

**(G)** Could the Wollaston coordinates have come from the measured position of a non-eruptive variable star? That is, might there be some ordinary variable star inside Wollaston's error box, and Wollaston measured this star at a time when it was brighter than HD 143707? The brightest star in or near Wollaston's error box (other than TCrB itself) is at 14th magnitude, so any such variable would have to have an amplitude of >7 mags. This could only be a rare high-amplitude Mira star. But the exhaustive measures of all stars in the area shows there to be no giant or even subgiant stars down to 17.0 magnitude.[34] Further, any such large-amplitude variable would have been easily discovered by many surveys long ago.

**(H)** Could the cataloged coordinates be some sort of an error, with the match to the T CrB position being by random happenstance. That is, the primary evidence for a T CrB eruption is that the coordinates match the nova's position, yet could it be that the printed numbers are like a typo or a clerical mixup such that the numbers are essentially random? If mistaken coordinates are drawn from any position on the sky, then the probability of Wollaston reporting the error box as a match to T CrB, or any one interesting target, is $8.5 \times 10^{-8}$. Wollaston connected the coordinates to Herschel's V 75, so maybe the mistakes are only in the printed numbers of arc-minutes on the entry line (i.e. 51 for RA and 29 for declination). But the probability that some typo or error on the two printed integers is 0.00028 that T CrB will be pointed at, so the error hypothesis is rejected at the 3.6-sigma confidence level.[35]

From this list of all possibilities for the source of the coordinates, only the T CrB hypothesis (A) and the HD hypothesis (B) are not already strongly refuted:

## Arguments for the HD hypothesis

**(pro-HD #1)** Wollaston's catalog explicitly identifies the double star as V 75 (i.e. HD 143707), and he repeats Herschel's description. With no further information, we can only take Wollaston at his word.

## Arguments against the HD hypothesis

**(con-HD #1)** Wollaston's catalog adds a long comment after the stereotyped quote from Herschel, and this quadruple star is certainly not HD 143707.[36] The inclusion of this description of some third star is an argument against the HD-hypothesis. But the identical argument is an equal argument against *all* hypotheses. As an equal argument against all hypotheses, this point becomes a no-argument for trying to decide the origin of the coordinates. Rather, Wollaston's inclusion of an added description of some unrelated quadruple star is pointing to the catalog entry being a unique special case that must have a complex backstory. Within the HD-hypothesis, there is no reason for any complex backstory.

**(con-HD #2)** Wollaston's quoted coordinates are wrong for HD 143707. Wollaston's quoted position is 0.39° off from that of HD 143707, and that is substantially larger than the expected uncertainty.[37] So in the HD hypothesis, Wollaston's 1789 catalog quoted coordinates should be within a few arc-minutes of HD 143707. But they are not. The catalog coordinates offset from HD 143707 (0.39°) is beyond any ordinary error of Wollaston. In particular, if the cataloged coordinates came from a measure of HD 143707 with a 0.18° one-sigma uncertainty (like for Herschel's non-specific accuracy), the probability is 0.000408 that the $0.22° \times 0.016°$ error box will include T CrB. But this small probability is not applicable for the specific case of V 75, because if Wollaston used the position derived from Herschel's description then his coordinates would be 0.075° in error, while if Wollaston used the coordinates from his own 1784 measure of HD 143707 then his error would be 0.0036°. In these realistic cases, the probability that the cataloged coordinates came from a measure of HD 143707 is $10^{-11}$ or smaller that the $0.22° \times 0.016°$ error box will include T CrB. That is, the HD-hypothesis requires a very improbable set of errors so that a measured position for HD 143707 results in the reported position being a small error box centered on T CrB. Such odds are so low that the hypothesis is unacceptable. This is a confident refutation of the HD-hypothesis.

## Arguments against the T CrB hypothesis

**(con-TCrB #1)** The added description of the star as a quadruple certainly does not fit T CrB.[38] This argument against the T CrB hypothesis is equally an argument against *all* hypotheses, and so is really a non-argument for distinguishing the source of the cataloged coordinates. The inclusion of this third-star data in the entry in Wollaston's catalog is really telling us that there is some complex backstory to the creation of the entry. The T CrB hypothesis has easy opportunity for a complex backstory, as Wollaston is seen to be struggling with his late-1787 measurement of the V 75 position (really for T CrB in eruption) being inconsistent with Herschel's and Wollaston's measured position for V 75 (i.e. with HD 143707).

## Arguments for the T CrB hypothesis

**(pro-TCrB #1)** The position of Wollaston's star is that of T CrB. This provides the confident and unique connection from Wollaston's catalog to T CrB.

**(pro-TCrB #2)** T CrB has a recurrence time scale of near 80 years. Modern light curves of the 10 known galactic recurrent novae show that the light curves are the same from eruption to eruption, and for T CrB in particular.[39] So we expect a prior eruption back in the year $1866 - 80 = 1786$ or so. This is what we see in Wollaston's catalog. So the timing of the eruption is a causal connection from Wollaston's catalog to T CrB. Given that recurrent novae have cycle times that are not constant, this timing argument is not strong. Nevertheless, Wollaston's catalog is presenting us with an otherwise-unknown star that appears at the exact position of of T CrB and at the time that a T CrB eruption is strongly expected.

**(pro-TCrB #3)** In Wollaston's catalog, all of Herschel's double stars (with no other catalog source) have comments that are only from Herschel's catalog. Nothing more. This universality has only one exception, and that is for the star labeled as "H V 75." This entry not only has coordinates from a different star, but extra comments from an unknown third star. This one catalog entry forces us into some sort of a complex backstory. If the cataloged coordinates are those of HD 143707, or any other normal star, there is no reason for any complexity. In the TCrB-hypothesis, the answer is easy. Wollaston measured coordinates of T CrB in outburst, and substituted these coordinates into his catalog, thinking that they were an improved position for V 75. Apparently, this was just before Wollaston's 1787 December 28 letter to Herschel where he asks Herschel for any positional information on double stars in his second catalog apparently in CrB, that are positioned with respect to oblique lines. So the unique complexity of the single catalog entry points to unusual circumstances, just like for an observation of T CrB in eruption.

## *Conclusions on the December 1787 nova*

Everyone expects that T CrB erupted around the year 1786, reaching a peak of 2.0 mag. So, a conclusion that Wollaston recorded the astrometry of T CrB in eruption is not an extraordinary claim. Rather, as an ordinary and expected event, the conclusion does not need any extraordinary evidence. We do not have extraordinary evidence of the 1787 eruption, as all we have are the close match of the coordinates of one entry in an old catalog, plus letters and charts from Wollaston detailing the circumstances. By themselves, the striking coincidence of both time and sky position with the expected T CrB eruption provide a strong argument that the coordinates actually are measurements from T CrB. Further, we do have evidences that allows us to strongly reject all other possible explanations at high confidence levels. With a T CrB eruption peaking around 1787 December 20 closely matching the available evidence, and all possible alternatives being strongly refuted, I judge that the 1787 eruption is proven.

# The nova of 1217 AD

The *Ursperger Chronicle*[40] (or *Chronicon Urspergensis*) is a typical medieval monastic chronicle, listing events on a year-by-year basis along with various commentaries, covering the time interval from 1126 to 1229. The entries up until 1225 were written by Burchard (*c.* 1177–1230), who was the Abbott of Ursberg Abbey (in southern Germany just west of Augsburg). In an entry for the year 1217 AD, Burchard says:

> "Eodem anno tempore autumpnali, hora vespertina post occasum solis in quadam stella in occidente visum est signum mirabile. Nam cum stella illa, posita versus austrum, aliquantulum declinans in occidentem, in directo sideris illius, quod vocant astrologi coronam Ariadnae, sicut nos ipsi annotavimus, antea erat parva et post ad parvitatem rediit, sed tunc maiori lumine refulsit, visusque est ab ea ascendere versus altitudinem firmamenti quidam radius valde clarus, quasi trabes magna et alta. Et hoc per multos dies, ut predictum est, tempore autumpnali sero visum est; post paulatim defecit et ad suam parvitatem stella rediit. Predicatores quoque his temporibus multa alia asserebant contigisse signa in celo et in terrs, quae longum esset enumerare et huic brevitati annectere."[41]

A translation of this central text, mostly based on the translation of C. M. Botley,[42] is:

"In the autumn season of the same year [1217 AD], in the early evening, a wonderful sign was seen in a certain star in the west. This star was located a little west of south, in what astrologers call Ariadne's Crown [Corona Borealis]. As we ourselves observed, it was originally a faint star that for a time shone with great light, and then returned to its original faintness. There was also a very bright ray reaching up the sky, like a large tall beam. This was seen for many days that autumn. The preachers also in these times asserted that many other signs had taken place in heaven and on Earth, which it would be too long to enumerate and to add to this brief account."

This is an eye-witness report written by the observer (Burchard).

This report is of a transient point-source star ("stella") appearing brightly in the constellation Corona Borealis with a duration of a week-to-a-month or so ("many days" within a season). This is exactly what we expect for a medieval report on an old eruption of T CrB. On the face of it, we have a confident T CrB nova event in the year 1217.[43]

Nevertheless, we need a closer consideration of the T CrB hypothesis and any alternative possibilities. As a stellar transient that appears bright to the eye lasting for a week or more, the only possibilities are T CrB, some other nova, a Galactic supernova, one of the planets, or a comet. The transient of 1217 cannot be a supernova, because any supernova bright enough to get to second magnitude or brighter will be sufficiently close so as to leave behind an 806-year-old supernova remnant that would be one of the brightest sources in the whole sky in radio and X-ray, but such is not seen anywhere in the vicinity of Corona Borealis. And the transient of 1217 cannot be some other nova, because any nova bright enough to get to second magnitude or brighter would be sufficiently close to Earth that the now-quiescent nova system would likely be one of the brightest cataclysmic variables in the sky, likely discovered a century ago by any of many surveys, but such is not seen anywhere in the vicinity of Corona Borealis.[44] (Well, other than the T CrB system itself.) The transient of 1217 cannot be any of the planets, because Corona Borealis is near 45° from the ecliptic. So the only possibilities to consider are that the 1217 transient was a bright comet or a prior eruption of T CrB.

To keep the pros and cons of the two hypotheses straight, the next four subsections will itemize and discuss the various evidences:

## Arguments for the comet hypothesis

Burchard's text describes a star, not a comet, yet nevertheless, I can present three arguments that the 1217 transient was a comet:

**(pro-comet #1)** Comets are more common than novae in records of the time.[45] So in the absence of any further information, the comet hypothesis is the default choice. This is a weak argument, because we have substantial extra information, hence the decision will be based on the other information.

**(pro-comet #2)** The chronicles of the Saint Stephani monastery[46] recorded a *possible* comet in the year 1217. If so, then at least one comet made it to naked eye brightness in 1217, and this could well be what the Abbott of Upsberg reported in the autumn of the year. This is a weak argument, because naked-eye comets are relatively frequent, and the existence of one *possible* comet in 1217 says little about what happened in Corona

Borealis in the autumn. This is also a weak argument because it is uncertain as to whether the Stephani chronicles are referring to a comet at all.[47]

**(pro-comet #3)** Burchard reports "There was also a very bright ray reaching up the sky, like a large tall beam," and this could be a description of a comet tail. If this "beam" is accepted as a comet tail, then this argument is decisive for the comet hypothesis. But this is not a strong argument because the descriptions of "beam" and "ray" have only a poor chance of referring to a comet tail: **(A)** Celestial transients that are certainly not comets were frequently reported to have "rays" and "beams." For the cases of historical transient reports that certainly do not involve comets, among those that I happen to have published on over the last four decades, the *majority* were claimed to have "rays" or "beams." These include Supernova 1006,[48] the Crab Supernova of 1054,[49] the tower over Ros Ela[50] in 1054, the Canterbury Event[51] of 1178, Supernova 1181,[52] Tycho's Supernova[53] of 1574, and Kepler's Supernova[54] of 1604. The point is that it is common for medieval accounts of non-cometary celestial transients to include descriptions of rays and beams **(B)** A wide variety of real phenomena can readily be described as displaying a "beam" or "ray," including auroral arcs, meteors, meteor trains, pillars, and noctilucent clouds. If such a phenomenon happened toward the northwest in the early evening, the chronicler can easily create a connection between the two rare events at nearly the same time and direction. For example, a sun pillar at sunset would appear in nearly the same sky direction as Corona Borealis, just a few minutes before the "stella" appears, so that Burchard would correctly report what he sees, with an implied association. **(C)** Modern commentators did not make the connection from Burchard's "ray" and "beam" to any comet. Rather, later interpreters ascribed the beam to be either an auroral arc or the zodiacal band.[55] So a cometary explanation for Burchard's "beam" is not obvious and is not preferred. The point is that a cometary connection is just one weak possibility out of many common possibilities.

## Arguments against the comet hypothesis

The comet hypothesis has three negative arguments, two of which are very strong:

**(con-comet #1)** The text says that the transient was a star ("stella") with no other qualifiers, and that can only mean a stellar point source, and not a comet. Dall'Olmo has exhaustively collected all medieval words and phrases for "comet," and those that include the word "stella" have a word appended clearly indicating a comet.[56] Typical comet phrases are "stella caudata," "stella cum facula," "stella mortis," and so on. Burchard says "stella" alone, so the transient is not a comet. This is a simple and strong argument against the comet-hypothesis.

**(con-comet #2)** The "stella" has no reported cometary motion despite being closely observed for a week or more. This is not a strong argument, as there is no reason to require that motion be reported by Burchard. Still, motion across the sky is a traditional criterion to distinguish between a comet and a nova, and the 1217 transient fails this comet criterion.

**(con-comet #3)** Throughout Europe and the world, throughout the medieval times and in all ages, comets are universally taken as the worst of bad omens, always pointing to widespread deaths and fall of empires.[57] For example, the widely seen comet of 1208 is reported in the Stephani annals and is immediately claimed to be an omen for regicide, wars, robberies, and fires.[58] Contrarily, the 1217 event is reported as a "wonderful sign" ("signum mirabile"), with very positive connotations. A medieval Abbott would never label a comet as a "wonderful sign." Further, the preachers make connections to other signs, and would likely have indicated if all those signs were uniformly evil/bad. That the 1217 event was a positive omen is a very strong argument against the 1217 transient being a comet.

## Arguments against the T CrB hypothesis

I can think of only one argument against the T CrB hypothesis, and that is weak:

**(con-TCrB #1)** The "ray" and the "beam" have no place with a TCrB-hypothesis. But this is a weak argument because the description of rays can enter into an account of a T CrB eruption for a variety of reasons that are common in medieval records. See the discussion for pro-comet #3 above.

## Arguments for the T CrB hypothesis

The T CrB hypothesis is very strong:

**(pro-TCrB #1)** The 1217 transient is a stellar point source visible for many days and can only be a nova. That is, the comet hypothesis is refuted (due to Burchard's label of "stella" and his attribution as a positive omen), the supernova hypothesis is not possible (due to the lack of any remnant in the area), and the planet hypothesis is impossible (due to the ecliptic latitude near 45°). This nova in 1217 cannot be from any nova other than T CrB (due to the lack of any other very bright cataclysmic variable in the area). So the only possibility is that this stellar transient in Corona Borealis was T CrB itself. This is a very strong argument.

**(pro-TCrB #2)** The position of the transient is in Corona Borealis. That is a good description for T CrB. The area on the sky in CrB is a very small fraction of the sky, seemingly very unlikely unless there is a causal connection. This provides a direct connection from the 1217 transient to T CrB.

**(pro-TCrB #3)** The transient was visible for "many days," so the 1217 event has a poorly defined duration somewhere from a week to a month or so. T CrB is visible to the naked eye for 7 days,[59] and so is consistent with the T CrB hypothesis. This is a consistency argument.

**(pro-TCrB #4)** The 1217 transient occurs 8x81 years before the 1866 T CrB eruption. T CrB has a cycle time of 80 years, so we should *expect* a T CrB eruption round about the year 1226. And this is what we see, hence making a connection from the 1217 event to T CrB. This is a weak argument, as RN cycle times are variable, and an extrapolation back in time by eight cycles might have too large an uncertainty to be useful.

## Conclusions on the 1217 transient

The TCrB-hypothesis is not an extraordinary claim, but rather we expect that T CrB should appear in medieval records just like we see in Burchard's chronicles. That is, we know that T CrB will appear as one of the brightest novae visible in the sky roughly once every three generations, and its eruptions have inevitably been discovered by many people over the last millennium.

The 1217 report by Burchard is what we expect for a surviving medieval report on one of T CrB's many past eruptions. Burchard saw a bright point source in Corona Borealis, with a duration of a week-to-a-month or so. Such can only be a previous eruption of T CrB.

The only alternative for the TCrB-hypothesis is that the 1217 transient was a comet. But this possibility is refuted by Burchard explicitly labeling it as "stella," which means a point source (and not labeling it with any of the terms for a comet). Further, Burchard calls it a "wonderful sign," and such a label is impossible for a comet. A skeptic can always claim that the "beam" is a comet tail instead of the various other likely cases. But the skeptic must also posit that Burchard mistakenly used the wrong term for a comet, that Burchard did not report the comet's motion, and that a medieval Abbott would call a comet as a "wonderful sign."

In all, the case for the 1217 transient being T CrB is strong, more than sufficiently strong to support a claim for an expected event. The only possible alternative has strong refutations. In the end, I conclude that Burchard saw an eruption of T CrB in the year 1217.

## Overall conclusions

The confidence in the existence of the eruptions in 1787 and 1217 is high, about as high as could be expected for any historical case. In both cases, we have definite connections from the observed transient back to T CrB itself. (In 1787, we have Wollaston's measured astrometry placing a non-moving star brighter than 7.8 mag, with the small error box right on top of T CrB. In 1217, we have Burchard's eyewitness report of a point-source star ("stella") that "shone with a great light" for "many days" as a very positive omen (a "wonderful sign" in Corona Borealis.) Unlike for human events, astronomical transients can have all possibilities enumerated, quantified, and tested. Both the 1787 and 1217 transients have all the non-TCrB possibilities strongly rejected. (In 1787, Wollaston had to have gotten the cataloged coordinates from somewhere, and all other possibilities have unacceptably small probabilities for producing a small error box on top of T CrB. In 1217, there are few possibilities to get a transient lasting "many days" that gets up to second magnitude, and Burchard's event cannot be a comet because he explicitly used the term for a point-source star and because he attributes the transient to be a wonderful omen.) Both reports fit the expectations for a T CrB eruption, and all other possibilities are eliminated. So we can be confident that the celestial transients of 1787 and 1217 were prior eruptions of the recurrent nova T CrB.

T CrB has four observed eruptions in the years 1217.8, 1787.9, 1866.4, and 1946.1, plus one more expected upcoming in 2024.4 ± 0.3. The recurrence timescales are 7 × 81.4, 78.5, 79.7, and likely 78.3 ± 0.3 years. With nine eruptions from 1217.8 to 1946.1, the average recurrence timescale is 80.9 years. I expect additional eruptions within a year or two of 1706, 1625, 1544, 1462, 1381, 1299, 1137, 1055, 974, 892, 811, 730, 648, 567, 485, 404, 323, 241, 160, and 78 AD. With T CrB being one of the all-time

brightest novae, each of these eruptions were likely discovered by many observers throughout the northern hemisphere.

## Acknowledgements

The author thanks R. Neuhauser and W. Steinicke for detailed general discussions, with such always being valuable. Further, R. H. van Gent and R. Ceragioli were helpful for seeking the source of Wollaston's quadruple star. M. Black, C. Drone-Silvers, and S. Prosser provided valuable assistance in tracking down rare manuscripts. And R. Webbink was instrumental for this whole enterprise by his construction of an exhaustive bibliography for T CrB.

## Notes on Contributor

Bradley E. Schaefer is a professor at the Louisiana State University, with extensive work throughout a long career on using historical observations for getting modern astrophysics results on novae and supernovae.

## Notes

Herschel collections. Royal Astronomical Society (RAS), *The Herschel Archive* (London: RAS, 2004).

21. A sure solution would come from Wollaston's logbooks and notebooks, but I have found nothing relevant. The archives of the Royal Society has a variety of Wollaston material at <https://catalogues.royalsociety.org/CalmView/Record.aspx?src=CalmView.Persons&id=NA3227>, while his commonplace book is mostly blank with only 12 pages of material that does not contain any observations. The Royal Astronomical Society only has twenty scattered and unanswered letters, mostly to W. Herschel, that contain no observtions. These letters show that Wollaston was asking Herschel for positions on his double stars from Herschel's second catalog, with no surviving reply from Herschel. Wollaston was promising errata for his catalog, and this finally appeared tacked onto the end of his 1800 paper with a catalog of circumpolar stars, but no mention is made of V 75. See F. Wollaston, *Fasciculus Astronomicus Containing Observations of the Northern Circumpolar Region* (London, 1800), p. 85.

22. The observer of the coordinates must be some astronomer experienced in astrometry, likely be from England (so as to communicate the position to Wollaston on a special case basis), and must be interested in double star astrometry. The total candidate pool is just W. Herschel and Wollaston himself. I have exhaustively searched Herschel's observing logbooks and so I know that he has nothing past the three night in 1882 and 1883, as reported in his double star catalog. So the observer of the coordinates can only be Wollaston personally. And Wollaston has shown (in multiple letters to Herschel) that he is interested in particular in the astrometry of double stars in CrB from Herschel's second catalog. So the case is strong for Wollaston himself being the source of the coordinates.

23. W. Herschel's observations of V 75 were on the nights of 1782 July 18, 1783 March 24, and 1783 August 25. RAS, *op. cit.* (Note 20), Journals number 4 and 5, folder Herschel W 2/1. Herschel collects all his observations of V 75 in RAS, *op. cit.* (Note 20), Herschel W 2/5 3, p. 112.

24. Herschel, *op. cit.* (Note 19), p. 109. Herschel reports the position of V 75 to be "E telefcopicis ε Coronæ borealis fequentibus" and "About 1-degree following ε, in a line parallel to θ and ε Coronæ; the preceding of three forming an arch." That is, V 75 is 1° away from ε CrB, on a oblique line towards the southeast, with that line being an extension of the oblique line from θ to ε CrB. This is illustrated in Figure 1. At that position, V 75 is westernmost of an arch of three stars (see Figure 1). This positional description actually returns a position quite close to the modern position of HD 143707. This is confirmation that Herschel's V 75 has the modern name HD 143707.

25. Herschel, *op. cit.* (1785, Note 18), p. 40.

26. Herschel, *loc. cit.* (Note 23). Further, Herschel himself scoured his notebooks for any double star observations after 1785, collecting them all into the third segment of his double star catalog, and nothing near V 75 is noted. See Herschel, *loc. cit.*, (1822, Note 18).

27. F. Wollaston, "On a Method of Describing the Relative Positions and Magnitudes of the Fixed Stars; Together with Some Astronomical Observations," *Philosophical Transactions of the Royal Society of London*, 74 (1784), 181–200.

28. Letters from F. Wollaston to J. Herschel, RAS Archives, W 1/13 W 182 (dated 1785 May 27) and W 1/13 W 183 (dated 1785 May 31).

29. Letter from F. Wollaston to J. Herschel, RAS Archives, W 1/13 W 186 (dated 1787 December 28).

30. Letter from F. Wollaston to J. Herschel, RAS Archives, W 1/13 W 190 (dated 1788 August 18).

31. F. Wollaston, "Observations of Miss Herschel's Comet, in August and September, 1786," *Philosophical Transactions of the Royal Society of London*, 77 (1887), 55–60.

32. With rare exceptions, all supernovae are less luminous than an absolute magnitude of −19.5. The visual extinction along the line of sight to T CrB is 0.18 mag. Wollaston's star will be near his limiting magnitude of 7.8. With the usual equation, this give a distance of 2.6 megaparsecs (8.6 million light-years). Making the supernova less luminous than the limit, or making the star appear brighter than its limit, makes for a somewhat closer distance, with the difference not being significant. Any such distance like 2.6 megaparsecs places the posited supernova very far outside our Milky Way galaxy (with radius of order 0.02 megaparsecs). All supernova must appear inside or near their original birth galaxy. At the distance of near 2.6 megaparsecs, the posited supernova was not born in our Milky Way, and any original host galaxy in that direction would appear very bright in the sky (approaching naked eye brightness). That no such galaxy appears along the line of sight in the direction towards CrB is proof that the birth galaxy does not exist, so there was no supernova reaching 7.8 mag in the direction of T CrB in any year.

33. The error box for Wollaston's coordinates are $0.22° \times 0.16°$, or $8.5 \times 10^{-8}$ of the sky. A total of 394 Galactic novae were discovered in the 133 years since 1890, of which only three (T CrB, YZ Ret, and CT Ser) are at Galactic latitudes >45°. The probability of a non-TCrB nova appearing anytime in the 18th century inside Wollaston's error box at a Galactic latitude of $+48.2°$ is of order $1.9 \times 10^{-7}$.

34. The *Gaia* database reports magnitudes, colors, and parallaxes for all stars brighter than 17th magnitude within 5 arc-minutes of T CrB, and none of these have an absolute magnitude more luminous than our Sun, nor the red colors of any type of long-period or semi-regular variable star. Gaia Collaboration *et al*., "The Gaia Mission," *Astronomy & Astrophysics*, 595 (2016), article ID A1, 36 pp. The *Gaia* data can be accessed at <https://gea.esac.esa.int/archive/>.

35. T CrB is inside the error box for the quoted coordinates only if the integers in the entry are 51 and 29, whereas in the error-hypothesis, the originals could have been any integer from 0 to 59. The probability that some original RA measure getting corrupted to 51 is 1/60, while the probability that some original declination report getting corrupted to 29 is 1/60. The joint probability for corrupting these two partial coordinates into the one pair of values that point to T CrB is $(1/60)^2$, or 0.00028. So, a random error in both of the printed arc-minutes so as to point to T CrB is rejected at the 3.6-sigma confidence level.

36. This is proven with the Gaia data, where there are no other stars to 13th mag within 0.1°. Gaia Collaboration *et al., loc. cit.* (Note 34).

37. For double stars with only Herschel descriptive positional data with no problems of ambiguity or vagueness, the positional errors have an average positional deviation of 0.18°. What is going on is that Wollaston's coordinates for these stars are taken solely from Herschel's verbal description of the star's placement with respect to nearby stars, and occasionally Herschel gives a poor description, but Herschel's words were usually good enough to place the double star with respect to bright cataloged stars to a typical accuracy of 0.18°. For the case of V 75, Herschel's verbal description actually produces a position that is 4.5 arc-minutes from HD 143707. Further, Wollaston personally measured the position of the star HD 143707, and his position is 0.22 arc-minutes from the modern position of HD 143707. Wollaston, *op. cit.* (Note 27), figure 2.

38. *Gaia* would reveal any star brighter than Wollaston's limit of 7.8 mag or Herschel's limit of near 11.8 mag, and none are shown from very-close separations to very-far separations. T CrB is not a quadruple system, nor even a double star. Gaia Collaboration *et al., loc. cit.* (Note 34).

39. Schaefer, *op. cit.* (2010, Note 1), pp. 285–307. Schaefer, *op. cit.* (2023, Note 1), Section 2.

40. Burchard, Conradus and G.H. Pertz, *Burchardi et Cuonradi Urspergensium Chronicon* (Hanover, 1874), pp. 105–6. The text can be accessed at <https://www.google.com/books/edition/Burchardi_et_Cuonradi_Urspergensium_Chro/tIAk56XortUC>. This text is in Latin, and is part of the vast *Ex Monumentis Germaniae* program of printing medieval chronicles.

41. Burchard, *op. cit.* (Note 40), pp. 105–6.

42. Botley, *op. cit.* (Note 14).

43. A key question is for the stated year of the transient reported to be in 1217, as an error in the year could allow other hypotheses. For this, Burchard was an eyewitness writing in 1225, just 8 years after the transient, so any large error in the year is very unlikely, and any small error in the year is unlikely. Detailed comparisons of Burchard's chronicled events with known history (by G. Pertz and by G. A. Loud) has revealed no significant chronological errors. I have examined the chronicled events within Burchard's adult life to always be in historical order, and the dated events to be correct. Out of these many events, there are two exceptions, with one date wrong by one year and another being wrong by one month, with both being still in the correct temporal order. Importantly, the 1217 transient is placed correctly between chronicled events (from the death of Pope Innocent III and the start of the Fifth Crusade followed by the death of Emperor Otto IV and the departure of the Crusader army from Germany) with the reported dates being correct and in the right order. So we can be confident that Burchard reported the year 1217 correctly.

44. Novae come to peak absolute magnitudes of $-7.45 \pm 1.33$, so in the direction of CrB with negligible extinction, the system must be $<780$ pc so as to appear brighter than V = 2 at peak. Such a distance would place the nova in the closest 2% out of all 402 known novae. Almost all quiescent novae have absolute magnitudes from $+7$ to $+2$ mags, which for distances $<780$ pc, makes the quiescent counterpart brighter than 11.5 to 16.5 mag. (RNe are the exceptions, where they have absolute magnitudes from $+2$ to $-4$ due to high accretion rates and red giant companion stars.) Such quiescent novae would all have been discovered (mostly from 50–100 years ago) and recognized in variability surveys, color surveys, and X-ray surveys covering the entire sky. All would have their close distances recognized by any of many methods, including with Gaia. No such nearby ex-novae is known for sky positions in-or-near to CrB. It is conceivable that some nova could have an absolute magnitude fainter than $+10$ and be overlooked in some corner of CrB. But such a case is very improbable, due to the high galactic latitude (0.5% of known novae), the close proximity (2% of all novae), and the rarity of the low quiescent luminosity ($<1\%$ of all novae).

45. Over the half-millenium from 1000–1500 AD, historical sources record 174 separate comet apparitions, 21 novae, and 3 supernovae. G.W. Kronk, *Cometography, a Catalog of Comets, Volume 1: Ancient - 1799* (Cambridge: Cambridge University Press, 1999), pp. 166–293. D.H. Clark and F.R. Stephenson, *The Historical Supernovae* (Oxford: Pergamon Press, 1977), pp. 48–49.

46. S. Stephani, *Annales S. Stephani Frisingensis*, (1448). Full Latin text in G. Waitz (ed.), *Monvmenta Germaniae historica, Inde ab anno Christi qvingentesimo vsqve ad annum millesimum et qvingentesimvm*, vol. 13 (Hannover, 1881), p. 56. The text is available on-line at <https://www.dmgh.de/mgh_ss_13/index.htm#page/56/mode/1up>.

47. The 'comet' in the Stephani chronicle for 1217 is too weak to be useable as an argument for the the comet-hypothesis: The chronicle's text has only four relevant words ("Stella comes visa est.," which might be translated as "A comet is visible."), but this is too sparse to be convincing of anything. The chronicle does use the phrase "stella comes" for what is likely a comet in the year 1208 (followed by a listing of dire omens), and so the phrase can apparently mean "comet" at least this one time. Nevertheless, a reasonable case can be made that the item, "stella comes" is not a comet at all. Indeed, in his exhaustive survey of medieval terminology for comets, Dall'Olmo does not recognize "stella comes" as any medieval terminology ever used for a comet. (U. Dall'Olmo, "Latin Terminology Relating to Aurorae, Comets, Meteors, and Novae," *Journal for the History of Astronomy*, 11 (1980), 10–27.) In medieval Latin, the word "comes" is a title of nobility in medieval Germany, roughly translated as 'Count'. This exact meaning for the word "comes" was certainly used in the Stephani annals for the year 1283, when "comes Emicho" was used as a title of nobility when discussing the historical personage Count Emicho. Further, the existence of a comet in 1217 is not mentioned by any other source, European or Asian. So the existence of a 1217 comet in the Stephani chronicles is either denied or ambiguous.